\documentclass[preprint2]{aastex} 

\begin{document} 
\title{ LOW ORDER p-MODES IN A BIPOLYTROPIC MODEL OF THE SUN}
\author{G. A. Pinz\'on \footnote{e-mail \textit{ gpinzon@ciencias.ciencias.unal.edu.co} } }  
\affil{Observatorio Astron\'omico Nacional,  Facultad de Ciencias\\
Universidad Nacional de Colombia, Bogot\'a, Colombia } 
\begin{abstract}  
\noindent \footnotesize{Based on the Solar Standard Model SSM we developed a solar model in hydrostatic equilibrium using two polytropes that describes both the \textit{radiative} and \textit{convective} zones of the solar interior. Then we apply small periodic and adiabatic perturbations on this bipolytropic model in order to obtain proper frequencies and proper functions. The frequencies obtained are in the \textbf{p-modes} range of low order  $0<l<20$ which agrees with the observational data, particularly with the so called five minutes solar oscillations.}
\end{abstract} 
\section{Introduction} 

\noindent Polytropic models have largely been used in the study of non radial adiabatic oscillations of a gaseous sphere [Cowling,1941; Kopal,1949; Scuflaire,1974; Tassoul,1980]. 

\noindent We have computed the first modes of a bipolytropic model whose indices $n_{2}=3.85$ and $n_{1}=1.5$ describes both the \textit{radiative} and \textit{convective} zones respectively of the solar interior. We used Cowling's approximation [Cowling,1941] which reduces the order of the system of differential equations to 2 (instead of 4). The radial part of the perturbation obeys equations (1) and (2) [Ledoux-Walraven,1958]:

\begin{eqnarray}
\frac{dv}{dr} &=&\left[ \frac{L_{l}^{2}}{\sigma ^{2}}-1\right] \frac{P^{%
\frac{2}{_{\Gamma _{1}}}}}{\rho }w  \label{ledoux} \\
\frac{dw}{dr} &=&\frac{1}{r^{2}}\left[ \sigma ^{2}-N^{2}\right] \frac{\rho }{%
P^{\frac{2}{\Gamma _{1}}}}v  \label{ledouxa}
\end{eqnarray}
where
\begin{equation}
v=r^{2}\delta rP^{\frac{1}{\Gamma _{1}}}
\end{equation}

\begin{equation}
w=\frac{P'}{P^{\frac{1}{\Gamma _{1}}}}
\end{equation}

\noindent are proper functions, $l$ is the degree of the spherical harmonic, $\Gamma _{1}$ is 
the adiabatic exponentent equal to  $\frac{5}{3}$, $\sigma$ is the angular
 frequency, $N$ is the \textit{Brunt-V\"{a}is\"{a}l\"{a}} frequency and  
$L_{l}$ is the \textit{Lamb} frequency.

\noindent The equations (1) and (2) and the boundary conditions
lead to a eigenvalue problem with eigenvalue $\sigma^{2}$ and this is the problem to solve.

\section{Polytropes} 
\noindent By a polytrope we understand a gas of particles with spherical symmetry, selfgravitating, in hydrostatic equilibrium and  with the state equation:

\begin{equation}
P=K\rho ^{\gamma }=K\rho ^{1+\frac{1}{n}}
\end{equation}

\noindent $K$ and $\gamma$ are parameters that depend only of  $n$ the polytropic index and the mass and radius of the configuration. The  polytrope theory
developed at the beginnings of the XIX century, can be used to know the dynamical structure 
of a star, that changes its thermodynamic state through a \textit{polytropic process} in which the specific heat remains constant.  This is  a true result for several regions of the Sun; In \textit{Fig.1} we can see that the pressure and the density 
are used to plot $\gamma=\frac{dLnP}{dLn\rho}$ $vs$ $x=r/R_{\odot}$ in the sophisticated SSM of Bahcall and Pinsonneault. Three regions clearly emerge. In each region the SSM output is approximated rather well by a straight line, indicating polytropic behavior [Hendry,1993]. Of these three regions the outermost one ($\gamma_{1}=5/3$) represents the \textit{convective} zone where heat transport is achieved by adiabatic convection. The inner regions constitute the \textit{radiative} zone ($\gamma_{2}=1.26$) where heat transport is achieved by electromagnetic waves. Our model based on this characteristic of the SSM allows to obtain a representation of the SSM in terms of two polytropes. 

\begin{center}
\begin{figure}[h]
\begin{center}
\includegraphics[angle=0,width=7cm]{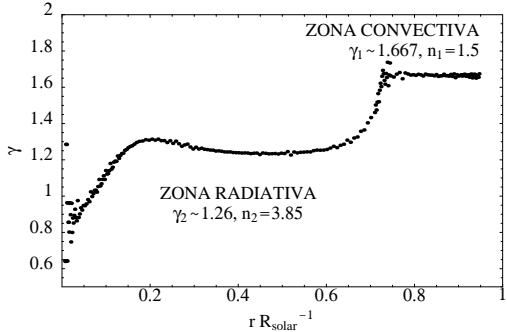}
\caption{\footnotesize {$\gamma$ at the solar interior due to SSM  [Bahcall et al.,1995]. Regions with  $\gamma$ constant are described by polytropic process. The bipolytropic  model is obtained if one assume two values for $\gamma$; 5/3 in the \textit{convective} zone, and 1.26 in the \textit{radiative} zone.}}
\end{center}
\end{figure}
\end{center}
 
\subsection{Two polytropes within the  Sun} \label{bozomath} 
\noindent Following Hendry [Hendry, 1993] we use $\xi$, $\theta$ as the variables in the \textbf{Lane-Emden} equation for 
the \textit{convective} zone with index $n_{1}$, and  $\eta$, $\phi$ as the variables for the \textit{radiative} zone with index $n_{2}$. The  parametric
 polytropes for the \textit{convective} and \textit{radiative} zones are 
respectively

\begin{eqnarray}
P &=&K_{1}\rho ^{\frac{5}{3}},\textrm{ with }\rho =\lambda _{1}\theta (\xi
)^{1.5}  \label{fit} \\
P &=&K_{2}\rho ^{1.26},\textrm{ with }\rho =\lambda _{2}\varphi (\eta )^{3.85}
\end{eqnarray}

\noindent The main challenge is to learn how to fit these two polytropes together. Since the physical quantities $P$, $\rho$ and $M$ are continuous across the interface (not $\theta$ or $\phi$), the variables $U$ and $V$ given by

\begin{equation}
U =\frac{\xi \theta ^{n}}{-\theta} 
\end{equation}
\begin{equation}
V =\frac{(n+1)\xi (-\theta)}{\theta}
\end{equation}
\noindent are be very useful [Chandrasekhar,1939].

\noindent Let us start by considering the \textit{convective} zone. We take $n_{1}=1.5$ thus $\xi_{1}=3.6538$ and $\theta'_{0}=-0.2033$. Though, this polytrope is not being used in the vecinity of $\xi=0$, it is possible to consider all of the solutions of the \textbf{Lane-Emden} equation for $n_{1}=1.5$. These may be generated beginnings  at $\xi_{1}$ with an arbitrary starting slope and integrating inwards. Solutions with starting slopes less negative than $\theta'_{0}$ are of particular interest (in the literature, they are referred as \textbf{M}-solutions) since these are the ones which intersect the polytrope that represents the \textit{radiative} zone. Three such solutions, translated into the $U$, $V$ variables, are shown in \textit{Fig.2}. Solutions with starting slopes more negative than $\theta'_{0}$ (\textbf{F}-solutions) do not intersect the \textit{radiative} polytrope and so do not need to be considered here. 

\noindent Knowing $\theta(\xi)$ and  $\phi(\eta)$ we can deduce the density and pressure curves $\rho(r)$ and $P(r)$. The above model yields a central density of $\rho_{c}=1.299\times10^{5}$ $Kg m^{-3}$ and a central pressure of $P_{c}=2.237\times10^{16}$ $N m^{-2}$. 

\begin{center}
\begin{figure}[h]
\begin{center}
\includegraphics[angle=0,width=7cm]{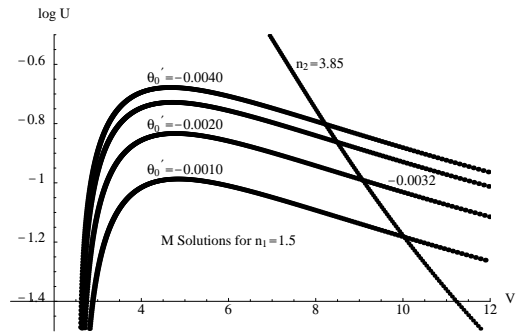}
\caption{\footnotesize {Solutions of the  \textbf{Lane-Emden} equation. The center of the Sun as at $U=3.0$, $V=0$ and its surface at $U=0$, $V=\infty$. 
We choose $\theta'_{0}=-0.0032$ as a \textbf{M}-Solution for the convective zone.}}
\end{center}
\end{figure}
\end{center}

\subsection{Characteristic Frequencies}

\noindent The characteristic \textit{Lamb} frequencies $L$ and \textit{Brunt-V\"{a}is\"{a}l\"{a}} frequency $N$ (in notation of Mullan-Ulrich,1981) associated with restoring forces of pressure and gravity in a polytrope of order $n$ can be calculated as functions of radius in terms of the Lane-Emden function $\theta$. For the definition of $\theta$, see, for example, Chandrasekhar (1939): in a perfect gas, $\theta$ equals the ratio of local temperature to central temperature. Thus, if we define

\begin{equation}
\omega _{g}=\frac{GM_{\odot }}{R_{\odot }^{3}}  \label{norm}
\end{equation}

\begin{center}
\begin{figure}[h]
\begin{center}
\includegraphics[angle=0,width=7cm]{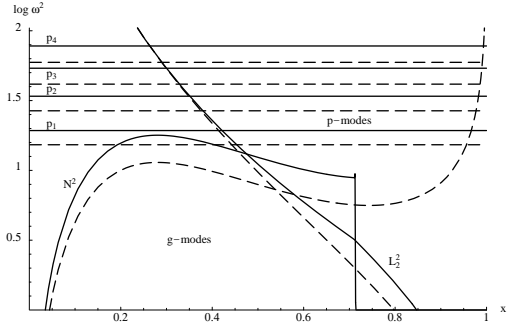}
\caption{\footnotesize {Propagation Diagram for the \textbf{bp} model (solid line) and for the polytropic index 3 model (dashed line).}}
\end{center}
\end{figure} 
\end{center}

\noindent where $R_{\odot }$ and $M_{\odot }$ are the sun mass and  the sun radius, and $G$
is the gravitational constant, we can write a normalized \textit{Brunt-V\"{a}is\"{a}l\"{a}} frequency as:

\begin{equation}
N_{n}^{2}\equiv \frac{N^{2}}{\omega _{g}}=\frac{(n-n_{0})}{(n_{0}+1)}\times
CC\times \frac{3}{\theta }(\frac{d\theta }{d\xi })^{2}  \label{ncuadnorm}
\end{equation}

\noindent where  $CC$ is the central condensation  of the polytrope, definided
as the ratio of central density to mean density. The parameter $n_{0}$ is 
the effective polytropic index associated with the pulsations [Mullan-Ulrich,1988], which 
are associated with the adiabatic exponent $\Gamma$ from:

\begin{equation}
\Gamma =1+\frac{1}{n_{0}}
\end{equation}

\noindent We supposed  $\Gamma =\frac{5}{3}.$

\noindent Similarly, the value of \textit{Lamb} associated with mode $l$ is given by

\begin{equation}
L_{\ln }\equiv \frac{L_{l}^{2}}{\omega _{g}}=3\times CC\times l(l+1)\times 
\frac{\theta }{\xi ^{2}}\times \frac{n_{0}+1}{n_{0}(n+1)}  \label{lambnorm}
\end{equation}

\noindent where $\xi $ is the Lane-Emden coordinate.

\section{NRO in a Bipolytropic Model}

\noindent Space oscillation properties of the solutions of equations (1) and (2) are related to the signs of the 
coefficients given in the second members of these equations. Space oscillations are allowed only in the regions
where these coefficients have opposite signs. The limits of the oscillations regions are defined by

\begin{equation}
\sigma ^{2}=\frac{l(l+1)\Gamma _{1}P}{\rho r^{2}}=L_{l}^{2}
\label{l}
\end{equation}

\begin{equation}
\sigma ^{2}=N^{2}  \label{n}
\end{equation}

\noindent In the $(x,\omega^{2})$ plane, these equations define two curves. 
In \textit{Fig.3} we have plotted them as a solid line for the 
bipolytropic model. Have been denoted \textit{p-modes} and \textit{g-modes} the regions of this plane corresponding to the conditions of position in the star and frequency, allowing spatial oscillations. These regions are characterized by the possibility of existence of progressive acoustic waves and progressive gravity waves respectively  
[Scuflaire,1974]. Thus we shall refer to these regions as the acoustic and the gravity regions. We have also plotted in the same figure the frequencies of the first \textit{p-modes} for bipolytropic model and $n=3$ model.

\noindent The equations (1) and (2) are very convenient for the analytical discussion, see for example [Unno et al.,1989], but for the numerical computations we use the more appropriate form

\begin{eqnarray}
\frac{dy}{dz}=\frac{l+1}{x}\left[ -y+\frac{l}{\omega ^{2}}z\right] + \frac{x}{%
\Gamma _{1}}\frac{GM_{\odot }\rho }{R_{\odot}P}\left( \frac{q}{x^{3}}y-z\right)  \label{sucuy}
\end{eqnarray}

\begin{eqnarray}
\frac{dz}{dx}=\frac{1}{x}\left[ \omega ^{2}y-lz\right] +{} R_{\odot}A\left( \frac{q}{%
x^{3}}y-z\right)  \label{sucuz}
\end{eqnarray}

\noindent where we have put

\begin{eqnarray}
\frac{r}{R_{\odot}} &=&x \\
\frac{m}{M_{\odot }} &=&q \\
\frac{\xi_{r}}{R_{\odot}} &=&x^{l-1}y \\
\frac{R_{\odot}P\prime }{GM_{\odot }\rho } &=&x^{l}z \\
\frac{R_{\odot}^{3}\sigma ^{2}}{GM_{\odot }} &=&\omega ^{2}
\end{eqnarray}

\noindent The regularity condition at the centre, first requires that  

\begin{equation}
\omega ^{2}y-lz=0  \label{center}
\end{equation}

\noindent and second, at the surface, the cancellation of the lagrangian perturbation of the pressure is written by

\begin{equation}
\frac{q}{x^{3}}y-z=0  \label{surface}
\end{equation}

\noindent In order to determine the  solution uniquely, we impose the normalizing condition
\begin{equation}
y=1  \label{norm}
\end{equation}
\noindent at the centre. With a trial value for $\omega ^{2}$ we integrate equations (16) and (17), with initial conditions (23) and (25) using Runge-Kutta method, with a  step size taken from paper of Christensen-Dalsgaard 
[Christensen-Dalsgaard et al.,1994]. Usually this solution does not satisfy equation (24) and a new integration is performed with another value of  $\omega ^{2}$. This procedure is repeated until equation (24) is satisfied, using a Newton-Raphson method to improve the value of  $\omega ^{2}$.

\begin{center}
\begin{figure}[h]
\begin{center}
\includegraphics[angle=0,width=8cm]{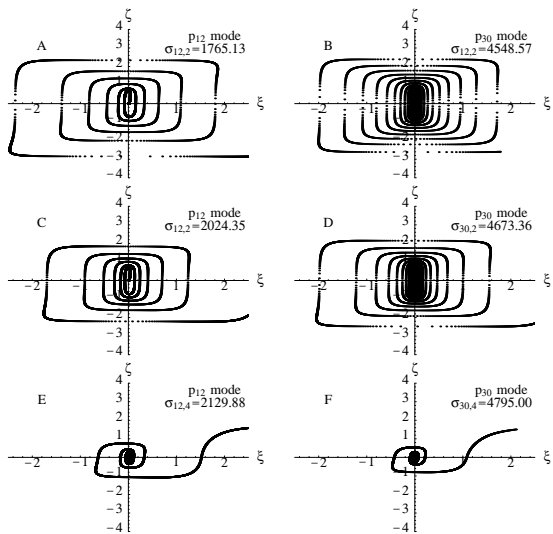}
\caption{{\footnotesize {\textbf{Some \textit{p-modes} for a polytropic
 models.} Figures \textbf{A} and \textbf{B} corresponds to the $n=3$ model. 
\textbf{C}, \textbf{D}, \textbf{E}, and  \textbf{F} are the same modes but in the bipolytropic model. The units of $\sigma_{nl}$ are $\mu$$Hz$.}}}
\end{center}
\end{figure}
\end{center}

\begin{center}
\begin{figure}[h]
\begin{center}
\includegraphics[angle=0,width=6cm]{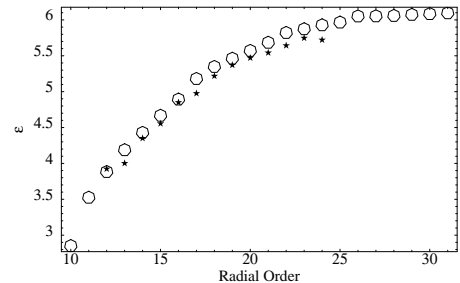}
\end{center}
\caption{\footnotesize {Comparison with GOLF data for $l=2$ (circle) and $l=4$ (star). The $\varepsilon$ value is given by 
\footnotesize{$\varepsilon$=$\frac{\sigma_{bp}-\sigma_{GOLF}}{\sigma_{bp}}\times100\%$}}}
\end{figure}
\end{center}

\noindent The radial displacement  $\delta(r)$ and the pressure perturbation 
$P'$ are periodic space functions; the variables $v(r)$ and $w(r)$ vary strongly from
the center to the surface, then it is impossible to plot them directly along the axes. However, the most appropiate functions \small{$\xi =\pm \log _{10}(1+\left| \frac{\delta(r)}{R_{\odot}}\right| )$ and
$\zeta =\pm \log _{10}(1+\left| \frac{R_{\odot}P'}{GM_{\odot }\rho }\right|)$}
, have been plotted in \textit{Fig.4}. Their signs are chosen according to the signs of the variables $\delta(r)$ and $P'$.

\section{Conclusion}
\noindent  Although we do not use an atmosphere model and the input physics is described by polytrope structure, the modes obtained 
from the bipolytropic model are close to the observacional data. We show in \textit{Fig.5} a comparison  with GOLF data.

\bigskip


\begin{thebibliography}{99}
\footnotesize{
\bibitem{}  Bahcall, M., Pinsonneault, M.:1995, Rev.Mod.Phys, \textbf{67},
 781
\bibitem{}  Chandrasekhar, S.:1939, \textit{An Introduction to the Study of
Stellar Structure}, Dover Publications, Chicago
\bibitem{}  Christensen-Dalsgaard J., Mullan, D.J.:1994, M.N.R.A.S.,
 \textbf{270}, 921
\bibitem{}  Cowling, T.G.:1941, M.N.R.A.S, \textbf{101}, 367
\bibitem{}  Hendry, A.:1993, Am.J.Phys, \textbf{61}, 10
\bibitem{}  Kopal, Z.1949,ApJ, \textbf{109},509
\bibitem{}  Ledoux, P., Walraven, T.:1958, Handbuch der Physik Vol LI, Springer Verlag, Berlin 
\bibitem{}  Mullan, D., Ulrich, R.:1988,ApJ, \textbf{331}, 1013
\bibitem{}  Scuflaire, R.:1974, A\&A, \textbf{36}, 107 
\bibitem{}  Tassoul, M.:1980,ApJss, \textbf{43}, 469
\bibitem{}  Unno, W., Osaki, Y., Ando, H., Saio, H., Shibahashi, H.:1989, 
\textit{Nonradial Oscillations of Stars}, University of Tokio Press, Tokio}

\end{thebibliography}
\end{document}